# A modified multiple-relaxation-time lattice Boltzmann model for convection-diffusion equation

Rongzong Huang, Huiying Wu[*]

*Key Laboratory for Power Machinery and Engineering of Ministry of Education,*

*School of Mechanical Engineering, Shanghai Jiao Tong University, Shanghai 200240, China*

*\* Corresponding author:* whysrj@sjtu.edu.cn

**Abstract:**

A modified lattice Boltzmann model with multiple relaxation times (MRT) for the convection-diffusion equation (CDE) is proposed. By modifying the relaxation matrix, as well as choosing the corresponding equilibrium distribution function properly, the present model can recover the CDE with anisotropic diffusion coefficient with no deviation term even when the velocity vector varies generally with space or time through the Chapman-Enskog analysis. This model is firstly validated by simulating the diffusion of a Gaussian hill, which demonstrates it can handle the anisotropic diffusion problem correctly. Then it is adopted to calculate the longitudinal dispersion coefficient of the Taylor-Aris dispersion. Numerical results show that the present model can further reduce the numerical error under the condition of non-zero velocity vector, especially when the dimensionless relaxation time is relatively large.

## 1. Introduction

The lattice Boltzmann equation (LBE) method is a powerful numerical technique for simulating fluid flows and complex physical processes in fluids [1-4]. Compared with the traditional computational fluid dynamics (CFD) method based on the macroscopic continuum equations, it has many notable merits, such as the mesoscopic kinetic background, easy boundary treatment and inherently parallelizable computation property. Initially, the LBE was derived from the lattice gas automata (LGA) to avoid statistic noise [5], and then it was proven to be a special discrete form (including the velocity discretization, time discretization, and space discretization) of the Boltzmann equation [6, 7]. In the past years, the LBE method has been used to simulate various flow problems (including the athermal flows [8, 9], thermal flows [10-12], compressible flows [13-15], porous media flows [16, 17], particulate flows [18, 19], etc). Also it has been designed to simulate the nonlinear problems described by nonlinear equations (including reaction-diffusion equation [20], wave equation [21, 22], etc.). As an important class of nonlinear equations, the convection-diffusion equation (CDE) has been solved by the LBE method extensively [23-39].

Most of the existing LBE models for CDE are commonly limited to the description of the isotropic diffusion problems [23-31]. Among these models, some are used to construct the thermal lattice Boltzmann models due to the isotropic thermal diffusivity [23-26]. In addition to the real CDE, the complex CDE is also solved by the LBE models with complex distribution function and relaxation time [27-29]. In the past years, some LBE models for CDE with anisotropic diffusion coefficient have been proposed [32-39]. Rasin et al. [32] proposed a two-dimensional multi-relaxation lattice kinetic method to handle the anisotropic diffusion problems. Following the similar way, Yoshida and Nagaoka [33] developed a three dimensional multiple-relaxation-time (MRT) LBE model for anisotropic diffusion problems. Ginzburg [34] presented a series of the equilibrium-type and link-type models to handle the anisotropic diffusion problems in two and three dimensions. And very recently, Ginzburg [35] summarized the alternative ways to realize anisotropic diffusion problems with a focus on their respective numerical diffusion.

Note that in most of the existing LBE models for CDE, deviation term exists in the corresponding macroscopic equations when the velocity vector varies with space or time (i.e., cannot reproduce the CDE precisely) through the Chapman-Enskog analysis (as will be analyzed in Section 2). Though this deviation term can be ignored under the condition of small Mach number $Ma$ (see Section 2 and Section 3), it still has significant influence on numerical error (as will be discussed in Section 3). Particularly, the deviation term disappears under the condition of zero velocity vector, or under the conditions of some special velocity vector profiles (such as $\mathbf{u}=const$, $\mathbf{u}=[u_x(y),0]$, $\mathbf{u}=[0,u_y(x)]$, etc.) by adopting appropriate equilibrium distribution function (i.e., including proper nonlinear terms of velocity). However, for most practical situations where the velocity vector varies generally with space or time, the deviation term still exists.

In the present paper, we propose a modified multiple-relaxation-time lattice Boltzmann model for CDE. By modifying the relaxation matrix, as well as choosing the corresponding equilibrium distribution function properly, it can reproduce the CDE with anisotropic diffusion coefficient with no deviation term even when the velocity vector varies generally with space or time through the Chapman-Enskog analysis. The remainder of the present paper is organized as follows. Section 2 presents the modified MRT model for CDE. Section 3 gives some typical numerical tests to validate the present modified MRT model, and a brief conclusion is drawn in Section 4. Appendix A gives a three-dimensional version of the present model with D3Q15 discrete velocity model.

## 2. Modified multiple-relaxation-time model for CDE

In this paper, we consider the following convection-diffusion equation (CDE) without source term:

$$\partial_t \phi + \nabla \cdot (\phi \mathbf{u}) = \nabla \cdot (\mathbf{D} \nabla \phi), \tag{1}$$

where $\phi$ is the conserved scalar variable (e.g. temperature $T$, etc.), $\mathbf{u}$ is the velocity vector which varies with space or time, and $\mathbf{D}$ is the symmetric matrix of diffusion coefficients. Here, the eigenvalues of $\mathbf{D}$ should be greater than zero to ensure the physical reality. Note that when $\mathbf{D} = \alpha \mathbf{I}$ ($\alpha$ is the diffusion coefficient and $\mathbf{I}$ is the unit matrix), the equation describes the isotropic diffusion problems. For the sake of simplicity, the two-dimensional situation will be studied in this paper, and its extension to the three-dimensional situation is straightforward (Appendix A gives a three-dimensional version of the present model).

*2.1. SRT models for CDE*

As a special discrete numerical method to solve CDE, the lattice Boltzmann equation describes the dynamic evolution process of the distribution function in discrete velocity space. In two-dimensional situations, the two-dimensional nine-velocity (D2Q9) discrete velocity model is commonly used. The nine discrete velocities are

$$\mathbf{e}_i = \begin{cases} c(0,0), & i = 0, \\ c(\cos[(i-1)\pi/2], \sin[(i-1)\pi/2]), & i = 1,2,3,4, \\ \sqrt{2}c(\cos[(2i-9)\pi/4], \sin[(2i-9)\pi/4]), & i = 5,6,7,8, \end{cases} \tag{2}$$

where $c = \delta_x/\delta_t$ is the lattice speed, $\delta_x$ is the lattice spacing, and $\delta_t$ is the time step. Note that $\delta_x$ and $\delta_t$, as well as all the other variables defined in this paper, are given to be dimensionless. The LBE with single relaxation time (SRT) can be expressed as follows

$$f_i(\mathbf{x}+\mathbf{e}_i\delta_t, t+\delta_t) = f_i(\mathbf{x},t) - \frac{1}{\tau}\left[f_i(\mathbf{x},t) - f_i^{eq}(\mathbf{x},t)\right], \quad i = 0 \sim 8, \tag{3}$$

which can be decomposed into two sub-steps, i.e., the collision process

$$f_i'(\mathbf{x},t) = f_i(\mathbf{x},t) - \frac{1}{\tau}\left[f_i(\mathbf{x},t) - f_i^{eq}(\mathbf{x},t)\right], \quad i = 0 \sim 8, \tag{4}$$

and the streaming process

$$f_i(\mathbf{x}+\mathbf{e}_i\delta_t, t+\delta_t) = f_i'(\mathbf{x},t), \quad i = 0 \sim 8, \tag{5}$$

where $f_i(\mathbf{x},t)$ is the distribution function in the direction $\mathbf{e}_i$, $\tau$ is the dimensionless relaxation time, and $f_i^{eq}(\mathbf{x},t)$ is the equilibrium distribution function which can be chosen as [23]

$$f_i^{eq} = \omega_i \phi \left[ 1 + \frac{\mathbf{e}_i \cdot \mathbf{u}}{c_s^2} + \frac{(\mathbf{e}_i \cdot \mathbf{u})^2}{2c_s^4} - \frac{\mathbf{u}^2}{2c_s^2} \right], \tag{6}$$

where $\omega_i$ is the weight coefficient in the direction $\mathbf{e}_i$ that is given by

$$\omega_i = \begin{cases} 4/9, & i = 0, \\ 1/9, & i = 1,2,3,4, \\ 1/36, & i = 5,6,7,8, \end{cases} \tag{7}$$

and $c_s = c/\sqrt{3}$ is the so-called sound speed. The conserved scalar variable $\phi$ is calculated as

$$\phi = \sum_{i=0}^{8} f_i. \tag{8}$$

Through the Chapman-Enskog analysis, the corresponding macroscopic equation is

$$\partial_t \phi + \nabla \cdot (\phi \mathbf{u}) = \nabla \cdot \left\{ \delta_t (\tau - 0.5)\left[\partial_t(\phi \mathbf{u}) + \nabla \cdot (\phi \mathbf{u}\mathbf{u}) + c_s^2 \nabla \phi \right] \right\}. \tag{9}$$

The diffusion coefficient matrix is given by $\mathbf{D} = c_s^2 \delta_t (\tau - 0.5)\mathbf{I}$. As compared with Eq. (1), there exists the deviation term $\nabla \cdot \left\{ \delta_t (\tau - 0.5)\left[\partial_t(\phi \mathbf{u}) + \nabla \cdot (\phi \mathbf{u}\mathbf{u})\right] \right\}$ in Eq. (9), which can be ignored under the condition of small Mach number (i.e., $\|\mathbf{u}\| \ll c_s$), or even disappear completely from Eq. (9) under the conditions of some special velocity vector profiles (such as $\mathbf{u} = const$, $\mathbf{u} = [u_x(y), 0]$, $\mathbf{u} = [0, u_y(x)]$, etc.) [31] due to the proper nonlinear terms of velocity in the equilibrium distribution function [34, 35].

*2.2. MRT models for CDE*

In order to recover the CDE precisely through the Chapman-Enskog analysis, the D2Q9 discrete velocity model with higher accuracy and more adjustable variables is necessary in the modified MRT model (note that D2Q5 with less computational cost is not suitable for the modified MRT model in the present paper). In MRT models based on D2Q9, the discrete velocities are given by Eq. (2), and the collision process is expressed as

$$f_i'(\mathbf{x},t) = f_i(\mathbf{x},t) - \Lambda_{ij}\left[f_j(\mathbf{x},t) - f_j^{eq}(\mathbf{x},t)\right], \quad i = 0 \sim 8, \tag{10}$$

where $\Lambda$ is the relaxation matrix in the velocity space. Generally, $\Lambda$ is a full matrix which makes the collision process not easy to be executed. However, if the collision process is carried out in the moment space, the relaxation matrix will be more concise. The moment $\mathbf{m}$ is calculated based on the distribution function $\mathbf{f} = (f_0, f_1, \cdots, f_8)^{\mathrm{T}}$ through a linear transformation

$$\mathbf{m} = \mathbf{M}\mathbf{f} = (m_0, m_1, \cdots, m_8)^{\mathrm{T}}, \tag{11}$$

where $\mathbf{M}$ is the transformation matrix which is constructed from the discrete velocities via the Gram-Schmidt orthogonalization procedure [40]. For the D2Q9 discrete velocity model, when the lattice speed satisfies $c \equiv 1$, $\mathbf{M}$ can be chosen as [40]

$$\mathbf{M} = \begin{bmatrix} 1 & 1 & 1 & 1 & 1 & 1 & 1 & 1 & 1 \\ -4 & -1 & -1 & -1 & -1 & 2 & 2 & 2 & 2 \\ 4 & -2 & -2 & -2 & -2 & 1 & 1 & 1 & 1 \\ 0 & 1 & 0 & -1 & 0 & 1 & -1 & -1 & 1 \\ 0 & -2 & 0 & 2 & 0 & 1 & -1 & -1 & 1 \\ 0 & 0 & 1 & 0 & -1 & 1 & 1 & -1 & -1 \\ 0 & 0 & -2 & 0 & 2 & 1 & 1 & -1 & -1 \\ 0 & 1 & -1 & 1 & -1 & 0 & 0 & 0 & 0 \\ 0 & 0 & 0 & 0 & 0 & 1 & -1 & 1 & -1 \end{bmatrix}. \tag{12}$$

Therefore, the collision process in the moment space can be expressed as

$$\mathbf{m}'(\mathbf{x},t) = \mathbf{m}(\mathbf{x},t) - \mathbf{S}\left[\mathbf{m}(\mathbf{x},t) - \mathbf{m}^{eq}(\mathbf{x},t)\right], \tag{13}$$

where $\mathbf{S} = \mathbf{M}\Lambda\mathbf{M}^{-1}$ is the relaxation matrix in the moment space, $\mathbf{m}^{eq}$ is the corresponding equilibrium function. Then the post-collision distribution function is obtained through the inverse linear transformation

$$\mathbf{f}'(\mathbf{x},t) = \mathbf{M}^{-1}\mathbf{m}'(\mathbf{x},t) = \left[f_0'(\mathbf{x},t), f_1'(\mathbf{x},t), \cdots, f_8'(\mathbf{x},t)\right]^{\mathrm{T}}. \tag{14}$$

After the collision process, the streaming process is executed in the velocity space, as Eq. (5) describes.

In order to get the CDE, the equilibrium function should be constructed firstly. Considering that only one conserved moment ($m_0$) exists in the model, there is no need to include higher-order velocity terms in the equilibrium function. So the equilibrium function is chosen as

$$m_0^{eq} = m_0 = \phi, \tag{15a}$$

$$m_1^{eq} = \frac{1}{4}\alpha_1\phi, \tag{15b}$$

$$m_2^{eq} = \frac{1}{4}\alpha_2\phi, \tag{15c}$$

$$m_3^{eq} = c_1\phi u_x, \qquad m_5^{eq} = c_1\phi u_y, \tag{15d}$$

$$m_4^{eq} = \frac{c_2}{2}\phi u_x, \qquad m_6^{eq} = \frac{c_2}{2}\phi u_y, \tag{15e}$$

$$m_7^{eq} = m_8^{eq} = 0. \tag{15f}$$

Here, in order to recover the convection term (see Eq. (32b)) and the diffusion term without deviation term (see Eq. (34)) properly, the parameters $c_1$ and $c_2$ in the equilibrium function should meet the following conditions

$$c_1 = 1, \quad c_1 + c_2/2 = 0, \tag{16}$$

and $\alpha_1$ and $\alpha_2$ are free parameters, which can be adjusted to achieve higher numerical performance for different problems. In the velocity space, the equilibrium distribution function can be calculated by the inverse linear transformation ($\mathbf{f}^{eq} = \mathbf{M}^{-1}\mathbf{m}^{eq}$), given as

$$f_i^{eq} = \begin{cases} \dfrac{1}{36}(-\alpha_1 + \alpha_2 + 4)\phi, & i = 0, \\ \dfrac{1}{144}\left(-\alpha_1 - 2\alpha_2 + 16 + 48\dfrac{\mathbf{e}_i \cdot \mathbf{u}}{c^2}\right)\phi, & i = 1,2,3,4, \\ \dfrac{1}{144}\left(2\alpha_1 + \alpha_2 + 16 + 12\dfrac{\mathbf{e}_i \cdot \mathbf{u}}{c^2}\right)\phi, & i = 5,6,7,8. \end{cases} \tag{17}$$

Note that Eq. (16) has been substituted into Eq. (17).

The Taylor series expansion for Eqs. (10) and (5) is

$$\mathbf{Df} + \frac{\delta_t}{2}\mathbf{D}^2\mathbf{f} + O(\delta_t^2) = -\frac{\mathbf{\Lambda}}{\delta_t}(\mathbf{f} - \mathbf{f}^{eq}), \tag{18}$$

where $\mathbf{D} = diag(D_0, D_1, \cdots, D_8)$, and $D_i = (\partial_t + \mathbf{e}_i \cdot \nabla)$. In the moment space, Eq. (18) is written as

$$\tilde{\mathbf{D}}\mathbf{m} + \frac{\delta_t}{2}\tilde{\mathbf{D}}^2\mathbf{m} + O(\delta_t^2) = -\frac{\mathbf{S}}{\delta_t}\left(\mathbf{m} - \mathbf{m}^{eq}\right), \tag{19}$$

where $\tilde{\mathbf{D}} = \mathbf{MDM}^{-1} = \mathbf{I}\partial_t + \mathbf{E}\cdot\nabla$, of which $\mathbf{I}$ is the unit matrix, $\mathbf{E} = (\mathbf{E}_x, \mathbf{E}_y)$, and

$$\mathbf{E}_x = \mathbf{M}\left[diag(e_{0x}, e_{1x}, \cdots, e_{8x})\right]\mathbf{M}^{-1}, \tag{20a}$$

$$\mathbf{E}_y = \mathbf{M}\left[diag(e_{0y}, e_{1y}, \cdots, e_{8y})\right]\mathbf{M}^{-1}. \tag{20b}$$

To deduce the CDE, the following Chapman-Enskog expansions are applied

$$\begin{aligned}\tilde{\mathbf{D}} &= \mathbf{I}\partial_t + \mathbf{E}\cdot\nabla = \mathbf{I}\left(\varepsilon\partial_{t1} + \varepsilon^2\partial_{t2}\right) + \varepsilon\mathbf{E}\cdot\nabla_1, \\ \mathbf{m} &= \mathbf{Mf} = \sum_{n=0}^{+\infty}\varepsilon^n\mathbf{m}^{(n)},\end{aligned} \tag{21}$$

where $\varepsilon$ is a small expansion parameter. Substituting Eq. (21) into Eq. (19), we can rewrite Eq. (19) in the consecutive orders of the parameter $\varepsilon$ as

$$\varepsilon^0 : \mathbf{m}^{(0)} = \mathbf{m}^{eq}, \tag{22a}$$

$$\varepsilon^1 : \left(\mathbf{I}\partial_{t1} + \mathbf{E}\cdot\nabla_1\right)\mathbf{m}^{(0)} = -\frac{\mathbf{S}}{\delta_t}\mathbf{m}^{(1)}, \tag{22b}$$

$$\varepsilon^2 : \partial_{t2}\mathbf{m}^{(0)} + \left(\mathbf{I}\partial_{t1} + \mathbf{E}\cdot\nabla_1\right)\left(\mathbf{I} - \frac{1}{2}\mathbf{S}\right)\mathbf{m}^{(1)} = -\frac{\mathbf{S}}{\delta_t}\mathbf{m}^{(2)}. \tag{22c}$$

To get the corresponding macroscopic equation, the relaxation matrix should be defined. Generally, the relaxation matrix in the moment space, $\mathbf{S}$, is a diagonal matrix, which means the collision processes of different moments are decoupled with each other. However, in order to handle the anisotropic diffusion problems, the collision processes of two moments relating to scalar flux should be expressed as [33]

$$m_3' = m_3 - s_{xx}\left(m_3 - m_3^{eq}\right) - s_{xy}\left(m_5 - m_5^{eq}\right), \tag{23a}$$

$$m_5' = m_5 - s_{yx}\left(m_3 - m_3^{eq}\right) - s_{yy}\left(m_5 - m_5^{eq}\right). \tag{23b}$$

Thus the relaxation matrix, $\mathbf{S}$, can be described as

$$\mathbf{S} = \begin{bmatrix} s_0 & 0 & 0 & 0 & 0 & 0 & 0 & 0 & 0 \\ 0 & s_1 & 0 & 0 & 0 & 0 & 0 & 0 & 0 \\ 0 & 0 & s_2 & 0 & 0 & 0 & 0 & 0 & 0 \\ 0 & 0 & 0 & s_{xx} & 0 & s_{xy} & 0 & 0 & 0 \\ 0 & 0 & 0 & 0 & s_4 & 0 & 0 & 0 & 0 \\ 0 & 0 & 0 & s_{yx} & 0 & s_{yy} & 0 & 0 & 0 \\ 0 & 0 & 0 & 0 & 0 & 0 & s_6 & 0 & 0 \\ 0 & 0 & 0 & 0 & 0 & 0 & 0 & s_7 & 0 \\ 0 & 0 & 0 & 0 & 0 & 0 & 0 & 0 & s_8 \end{bmatrix}. \tag{24}$$

Introducing $\mathbf{A} = \begin{pmatrix} s_{xx} & s_{xy} \\ s_{yx} & s_{yy} \end{pmatrix}$, the equations for the conserved moment, $m_0$, in Eqs. (22a) ~ (22c) can be written as

$$\varepsilon^0 : m_0^{(0)} = m_0^{eq}, \tag{25a}$$

$$\varepsilon^1 : \partial_{t1} m_0^{(0)} + \partial_{x1} m_3^{(0)} + \partial_{y1} m_5^{(0)} = -\frac{s_0}{\delta_t} m_0^{(1)}, \tag{25b}$$

$$\varepsilon^2 : \partial_{t2} m_0^{(0)} + \partial_{t1}\left[\left(1 - \frac{s_0}{2}\right) m_0^{(1)}\right] + \nabla_1 \cdot \left[\left(\mathbf{I} - \frac{\mathbf{A}}{2}\right)\begin{pmatrix} m_3^{(1)} \\ m_5^{(1)} \end{pmatrix}\right] = -\frac{s_0}{\delta_t} m_0^{(2)}. \tag{25c}$$

Eq. (25a) indicates

$$\varepsilon^0 : m_0^{(n)} = 0, \quad \forall n \geq 1. \tag{26a}$$

Eqs. (25b) and (25c) can be simplified as

$$\varepsilon^1 : \partial_{t1}\phi + \partial_{x1}(\phi u_x) + \partial_{y1}(\phi u_y) = 0, \tag{26b}$$

$$\varepsilon^2 : \partial_{t2}\phi + \nabla_1 \cdot \left[\left(\mathbf{I} - \frac{\mathbf{A}}{2}\right)\begin{pmatrix} m_3^{(1)} \\ m_5^{(1)} \end{pmatrix}\right] = 0. \tag{26c}$$

In order to further simplify Eq. (26c), we combine the fourth equation with sixth equation in Eq. (22b), and then have

$$-\frac{\mathbf{A}}{\delta_t}\begin{pmatrix} m_3^{(1)} \\ m_5^{(1)} \end{pmatrix} = \begin{bmatrix} \partial_{t1} m_3^{(0)} + \partial_{x1}\left(\frac{2}{3} m_0^{(0)} + \frac{1}{6} m_1^{(0)} + \frac{1}{2} m_7^{(0)}\right) + \partial_{y1} m_8^{(0)} \\ \partial_{t1} m_5^{(0)} + \partial_{x1} m_8^{(0)} + \partial_{y1}\left(\frac{2}{3} m_0^{(0)} + \frac{1}{6} m_1^{(0)} - \frac{1}{2} m_7^{(0)}\right) \end{bmatrix}$$
$$= \partial_{t1}(\phi \mathbf{u}) + \left(\frac{2}{3} + \frac{1}{24}\alpha_1\right)\nabla_1 \phi. \tag{27}$$

Thus Eq. (26c) can be written as

$$\varepsilon^2 : \partial_{t2}\phi = \nabla_1 \cdot \left\{ \delta_t \left( \mathbf{I} - \frac{\mathbf{A}}{2} \right) \mathbf{A}^{-1} \left[ \partial_{t1}(\phi \mathbf{u}) + \left( \frac{2}{3} + \frac{1}{24}\alpha_1 \right) \nabla_1 \phi \right] \right\}. \tag{28}$$

Combining Eq. (26b) with Eq. (28), we can get

$$\partial_t \phi + \nabla \cdot (\phi \mathbf{u}) = \nabla \cdot \left\{ \delta_t \left( \mathbf{I} - \frac{\mathbf{A}}{2} \right) \mathbf{A}^{-1} \left[ \partial_t (\phi \mathbf{u}) + \left( \frac{2}{3} + \frac{1}{24}\alpha_1 \right) \nabla \phi \right] \right\}. \tag{29}$$

The diffusion coefficient matrix is given by $\mathbf{D} = \left( \frac{2}{3} + \frac{1}{24}\alpha_1 \right) \delta_t \left( \mathbf{I} - \frac{\mathbf{A}}{2} \right) \mathbf{A}^{-1}$. Obviously, this MRT model can handle the anisotropic diffusion problems. However, there exists the deviation term $\nabla \cdot \left[ \delta_t (\mathbf{I} - \mathbf{A}/2) \mathbf{A}^{-1} \partial_t (\phi \mathbf{u}) \right]$ in Eq. (29) as compared with Eq. (1). Note that this deviation term disappears only under the condition of zero velocity vector because no proper nonlinear terms of velocity are included in the corresponding equilibrium function.

In order to eliminate the deviation term from Eq. (29), as well as inspired by the idea of Zheng et al. [12], we modify further the collision processes of two moments relating to scalar flux as

$$m_3' = m_3 - s_{xx}(m_3 - m_3^{eq}) - \left( \frac{s_{xx}}{2} - 1 \right) s_4 (m_4 - m_4^{eq}) - s_{xy}(m_5 - m_5^{eq}) - \frac{s_{xy}}{2} s_6 (m_6 - m_6^{eq}), \tag{30a}$$

$$m_5' = m_5 - s_{yx}(m_3 - m_3^{eq}) - \frac{s_{yx}}{2} s_4 (m_4 - m_4^{eq}) - s_{yy}(m_5 - m_5^{eq}) - \left( \frac{s_{yy}}{2} - 1 \right) s_6 (m_6 - m_6^{eq}). \tag{30b}$$

Therefore, the modified relaxation matrix, $\mathbf{S}$, is

$$\mathbf{S} = \begin{bmatrix} s_0 & 0 & 0 & 0 & 0 & 0 & 0 & 0 & 0 \\ 0 & s_1 & 0 & 0 & 0 & 0 & 0 & 0 & 0 \\ 0 & 0 & s_2 & 0 & 0 & 0 & 0 & 0 & 0 \\ 0 & 0 & 0 & s_{xx} & \left( \frac{s_{xx}}{2} - 1 \right) s_4 & s_{xy} & \frac{s_{xy}}{2} s_6 & 0 & 0 \\ 0 & 0 & 0 & 0 & s_4 & 0 & 0 & 0 & 0 \\ 0 & 0 & 0 & s_{yx} & \frac{s_{yx}}{2} s_4 & s_{yy} & \left( \frac{s_{yy}}{2} - 1 \right) s_6 & 0 & 0 \\ 0 & 0 & 0 & 0 & 0 & 0 & s_6 & 0 & 0 \\ 0 & 0 & 0 & 0 & 0 & 0 & 0 & s_7 & 0 \\ 0 & 0 & 0 & 0 & 0 & 0 & 0 & 0 & s_8 \end{bmatrix}, \tag{31}$$

and accordingly, the equations for the conserved moment, $m_0$, in Eqs. (22a) ~ (22c) are changed to

$$\varepsilon^0 : m_0^{(0)} = m_0^{eq}, \tag{32a}$$

$$\varepsilon^1 : \partial_{t1} m_0^{(0)} + \partial_{x1} m_3^{(0)} + \partial_{y1} m_5^{(0)} = -\frac{s_0}{\delta_t} m_0^{(1)}, \tag{32b}$$

$$\varepsilon^2 : \partial_{t2} m_0^{(0)} + \partial_{t1}\left[\left(1-\frac{s_0}{2}\right)m_0^{(1)}\right] + \nabla_1 \cdot \left[\left(\mathbf{I}-\frac{\mathbf{A}}{2}\right)\begin{pmatrix} m_3^{(1)} + \frac{s_4}{2} m_4^{(1)} \\ m_5^{(1)} + \frac{s_6}{2} m_6^{(1)} \end{pmatrix}\right] = -\frac{s_0}{\delta_t} m_0^{(2)}. \tag{32c}$$

Note that as compared with Eqs. (25a) ~ (25c), only the second-order equation (32c) changes. Similarly, Eq. (32a) indicates

$$\varepsilon^0 : m_0^{(n)} = 0, \quad \forall n \geq 1. \tag{33a}$$

Eqs. (32b) and (32c) can be simplified as

$$\varepsilon^1 : \partial_{t1}\phi + \partial_{x1}(\phi u_x) + \partial_{y1}(\phi u_y) = 0, \tag{33b}$$

$$\varepsilon^2 : \partial_{t2}\phi + \nabla_1 \cdot \left[\left(\mathbf{I}-\frac{\mathbf{A}}{2}\right)\begin{pmatrix} m_3^{(1)} + \frac{s_4}{2} m_4^{(1)} \\ m_5^{(1)} + \frac{s_6}{2} m_6^{(1)} \end{pmatrix}\right] = 0. \tag{33c}$$

In order to simplify Eq. (33c) further, we add the fifth equation to the fourth equation, the seventh equation to the sixth equation in Eq. (22b), and then combine them together. Finally we can obtain the following equation

$$-\frac{\mathbf{A}}{\delta_t}\begin{pmatrix} m_3^{(1)} + \frac{s_4}{2} m_4^{(1)} \\ m_5^{(1)} + \frac{s_6}{2} m_6^{(1)} \end{pmatrix} = \begin{bmatrix} \partial_{t1}\left(m_3^{(0)} + m_4^{(0)}\right) + \partial_{x1}\left(\frac{2}{3}m_0^{(0)} + \frac{1}{2}m_1^{(0)} + \frac{1}{3}m_2^{(0)} - \frac{1}{2}m_7^{(0)}\right) + 2\partial_{y1} m_8^{(0)} \\ \partial_{t1}\left(m_5^{(0)} + m_6^{(0)}\right) + 2\partial_{x1} m_8^{(0)} + \partial_{y1}\left(\frac{2}{3}m_0^{(0)} + \frac{1}{2}m_1^{(0)} + \frac{1}{3}m_2^{(0)} + \frac{1}{2}m_7^{(0)}\right) \end{bmatrix}$$
$$= \left(\frac{2}{3} + \frac{1}{8}\alpha_1 + \frac{1}{12}\alpha_2\right)\nabla_1 \phi. \tag{34}$$

Thus Eq. (33c) can be written as

$$\varepsilon^2 : \partial_{t2}\phi = \nabla_1 \cdot \left[\left(\frac{2}{3} + \frac{1}{8}\alpha_1 + \frac{1}{12}\alpha_2\right)\delta_t\left(\mathbf{I}-\frac{\mathbf{A}}{2}\right)\mathbf{A}^{-1}\nabla_1 \phi\right]. \tag{35}$$

Combining Eq. (33b) with Eq. (35), we can get

$$\partial_t \phi + \nabla \cdot (\phi \mathbf{u}) = \nabla \cdot \left[\left(\frac{2}{3} + \frac{1}{8}\alpha_1 + \frac{1}{12}\alpha_2\right)\delta_t\left(\mathbf{I}-\frac{\mathbf{A}}{2}\right)\mathbf{A}^{-1}\nabla \phi\right]. \tag{36}$$

The diffusion coefficient matrix is given by $\mathbf{D} = \left(\frac{2}{3} + \frac{1}{8}\alpha_1 + \frac{1}{12}\alpha_2\right)\delta_t\left(\mathbf{I}-\frac{\mathbf{A}}{2}\right)\mathbf{A}^{-1}$. As compared with Eq. (1), no deviation term exists in Eq. (36), while in Eqs. (9) and (29), deviation term exists.

Note that there is a similar part, $\nabla \cdot [\mathbf{k}\partial_t(\phi\mathbf{u})]$ ($\mathbf{k}$ is a constant factor), in the deviation term of Eqs. (9) and (29). From the Chapman-Enskog analysis, it can be found that (see Eq. (27)) this part is induced by the influence of the convection term (which is recovered in $\varepsilon^1$ order, i.e., $\nabla \cdot (\phi\mathbf{u}) \propto \varepsilon$) on the diffusion term (which is recovered in $\varepsilon^2$ order, i.e., $\nabla \cdot (\mathbf{D}\nabla\phi) \propto \varepsilon^2$). So for general situation, this deviation term cannot be eliminated by modifying the equilibrium distribution function merely, or adopting a MRT model simply. However, by modifying the relaxation matrix (as Eq. (31) shows), as well as choosing the corresponding equilibrium function properly (as Eq. (15) shows), this deviation term can disappear from the corresponding macroscopic equation through the Chapman-Enskog analysis.

## 3. Numerical tests

Numerical tests are carried out in this section to validate the modified MRT model proposed above. For the modified model (Eq. (31)), the free parameters in the equilibrium function are chosen as $\alpha_1=-8$, $\alpha_2=8$, so that Eq. (15) can be written as

$$\mathbf{m}^{eq} = \phi(1,-2,2,u_x,-u_x,u_y,-u_y,0,0)^T, \tag{37}$$

this modified model is labeled by MRT-A. For the unmodified model (Eq. (24)), three kinds of equilibrium functions are used. The first one is that described by Eq. (6) which can be transformed in the moment space as

$$\mathbf{m}^{eq} = \phi(1,-2+3\mathbf{u}^2,1-3\mathbf{u}^2,u_x,-u_x,u_y,-u_y,u_x^2-u_y^2,u_xu_y)^T, \tag{38}$$

this unmodified model is labeled by MRT-B1. The second one is that has the same mass and convection weights with Eq. (37) but with the proper nonlinear terms of velocity (aimed at removing the deviation term in Eq. (29) under the conditions of some special velocity vector profiles (such as $\mathbf{u}=const$, $\mathbf{u}=[u_x(y),0]$, $\mathbf{u}=[0,u_y(x)]$, etc. ) [34, 35], just like that in Eq. (9)), which is given as

$$\mathbf{m}^{eq} = \phi(1,-2+3\mathbf{u}^2,2-3\mathbf{u}^2,u_x,-u_x,u_y,-u_y,u_x^2-u_y^2,u_xu_y)^T, \tag{39}$$

this unmodified model is labeled by MRT-B2. The third one is that described by Eq. (37), and this unmodified model is labeled by MRT-B3. With these equilibrium functions, the diffusion coefficient matrices are all expressed as $\mathbf{D} = \delta_t(\mathbf{I}-\mathbf{A}/2)\mathbf{A}^{-1}/3$, and therefore we can make sure that the relaxation factors ($s_0 \sim s_8$, $s_{xx}$, $s_{xy}$, $s_{yx}$, $s_{yy}$) are consistent for all the four models.

## 3.1. Diffusion of a Gaussian hill

The initial-value problem for Eq. (1) with constant velocity vector can be solved analytically. Provided that the initial value of the scalar variable obeys the Gaussian distribution (also known as the Gaussian hill) as follows

$$\phi(\mathbf{x},0) = \frac{\phi_0}{2\pi\sigma_0^2}\exp\left(-\frac{\mathbf{x}^2}{2\sigma_0^2}\right), \tag{40}$$

where $\phi_0$ is the total concentration and $\sigma_0^2$ is the initial variance, the analytical solution of Eq. (1) can be expressed as [33]

$$\phi(\mathbf{x},t) = \frac{\phi_0}{2\pi\sqrt{\|\boldsymbol{\sigma}_t\|}}\exp\left\{-\frac{1}{2}\boldsymbol{\sigma}_t^{-1}:\left[(\mathbf{x}-\mathbf{u}t)(\mathbf{x}-\mathbf{u}t)\right]\right\}, \tag{41}$$

where $\boldsymbol{\sigma}_t = \sigma_0^2\mathbf{I} + 2t\mathbf{D}$, $\|\boldsymbol{\sigma}_t\|$ and $\boldsymbol{\sigma}_t^{-1}$ are the determinant value and inverse matrix of $\boldsymbol{\sigma}_t$, respectively. In the numerical test, the computation domain is $x\in[-1,1]$ and $y\in[-1,1]$, which is divided by $512\times512$ lattice grids. The total concentration is $\phi_0 = 2\pi\sigma_0^2$ with $\sigma_0 = 0.05$ (here, $\sigma_0$ is small enough to adopt periodic boundary condition). Three types of diffusion coefficient matrices are considered

$$\mathbf{D}\times10^4 = \begin{pmatrix}2 & 0\\0 & 2\end{pmatrix}, \begin{pmatrix}1 & 0\\0 & 4\end{pmatrix}, \begin{pmatrix}1 & 1\\1 & 4\end{pmatrix}, \tag{42}$$

which represent the isotropic diffusion, diagonal anisotropic diffusion and full anisotropic diffusion problems, respectively.

Firstly, we consider the pure diffusion problems (i.e., $\mathbf{u}\equiv\mathbf{0}$). Fig. 1 shows the comparisons of the numerical results of MRT-A, MRT-B1, MRT-B2 and MRT-B3 with the analytical solutions for the contour lines of the scalar variable distribution at time $t_m$ when $\phi_{\max}(\mathbf{x},t_m) = 0.5\phi_{\max}(\mathbf{x},0)$. It is seen that the numerical results of MRT-A, MRT-B1, MRT-B2 and MRT-B3 are in good agreement with the analytical solutions. Figs. 1 (b) and (c) demonstrate that all the MRT models can handle the anisotropic diffusion problems precisely. Table 1 gives the average and maximum errors between the numerical results and the analytical solutions at time $t_m$. The average error $Er_2$ and maximum error $Er_\infty$ are defined by

$$Er_2 = \sqrt{\frac{1}{N}\sum_{x,y}(\phi_{\text{numeric}}-\phi_{\text{analytic}})^2}, \qquad Er_\infty = \max\{|\phi_{\text{numeric}}-\phi_{\text{analytic}}|\}, \tag{43}$$

where $N$ is the total number of the grid points. It is found that the errors of MRT-B1, MRT-B2 and MRT-B3 are smaller than those of MRT-A at $\mathbf{u} \equiv \mathbf{0}$. This phenomenon may be attributed to the following two reasons: (1) under the condition of zero velocity vector, the deviation term in the corresponding macroscopic equation of MRT-B1, MRT-B2 and MRT-B3 disappears and has no effects on the numerical results, as mentioned in Section 2; (2) a more concise relaxation matrix is adopted in MRT-B1, MRT-B2 and MRT-B3 than in MRT-A.

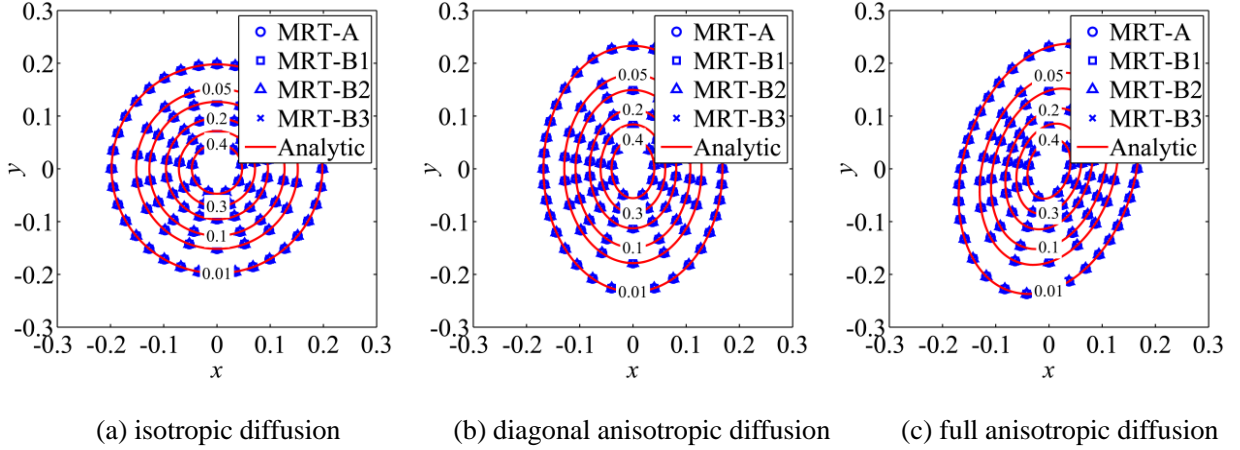

(a) isotropic diffusion  (b) diagonal anisotropic diffusion  (c) full anisotropic diffusion

**Fig. 1.** Contour lines of the scalar variable distribution at time $t_m$ and $\mathbf{u} \equiv \mathbf{0}$ based on the numerical results of MRT-A, MRT-B1, MRT-B2, MRT-B3 and the analytical solutions for (a) isotropic diffusion problem, (b) diagonal anisotropic diffusion problem, and (c) full anisotropic diffusion problem.

**Table 1**

Average errors $Er_2$ and maximum errors $Er_\infty$ between the numerical results of MRT-A, MRT-B1, MRT-B2 and MRT-B3 and the analytical solutions at time $t_m$ and $\mathbf{u} \equiv \mathbf{0}$.

| Diffusion coefficient matrix | MRT-A | | MRT-B1 | | MRT-B2 | | MRT-B3 | |
|---|---|---|---|---|---|---|---|---|
| | $Er_2$ | $Er_\infty$ | $Er_2$ | $Er_\infty$ | $Er_2$ | $Er_\infty$ | $Er_2$ | $Er_\infty$ |
| Isotropic | 2.242e-5 | 5.585e-4 | 1.595e-5 | 3.942e-4 | 2.017e-5 | 5.049e-4 | 2.017e-5 | 5.049e-4 |
| Diagonal anisotropic | 2.252e-5 | 5.487e-4 | 1.652e-5 | 3.966e-4 | 2.029e-5 | 4.960e-4 | 2.029e-5 | 4.960e-4 |
| Full anisotropic | 2.263e-5 | 5.327e-4 | 1.547e-5 | 3.658e-4 | 2.030e-5 | 4.876e-4 | 2.030e-5 | 4.876e-4 |

Now we consider the convection-diffusion problems (i.e., $\mathbf{u} = const \neq \mathbf{0}$), more general cases than the above pure diffusion problems. Here the velocity vector is set as $\mathbf{u} \equiv (0.05, 0.05)$. Table 2 gives the average and maximum errors between the numerical results and the analytical solutions at time $t_m$. It is found that the errors of MRT-B3 become much larger than those of MRT-A, MRT-B1 and MRT-B2 at $\mathbf{u} \equiv (0.05, 0.05)$. As compared with the corresponding values in Table 1, the errors of MRT-B3 in Table 2 increase remarkably, while the errors of MRT-A, MRT-B1 and MRT-B2 keep in the same level. This indicates that when the velocity vector is not equal to zero, the numerical error of MRT-B3 induced by the deviation term may become dominating, while the numerical errors of MRT-A, MRT-B1 and MRT-B2 are still relatively small because no deviation term exists in MRT-A, MRT-B1 and MRT-B2 when $\mathbf{u} = const \neq \mathbf{0}$.

**Table 2**

Average errors $Er_2$ and maximum errors $Er_\infty$ between the numerical results of MRT-A, MRT-B1, MRT-B2 and MRT-B3 and the analytical solutions at time $t_m$ and $\mathbf{u} = (0.05, 0.05)$.

| Diffusion coefficient matrix | MRT-A | | MRT-B1 | | MRT-B2 | | MRT-B3 | |
|---|---|---|---|---|---|---|---|---|
| | $Er_2$ | $Er_\infty$ | $Er_2$ | $Er_\infty$ | $Er_2$ | $Er_\infty$ | $Er_2$ | $Er_\infty$ |
| Isotropic | 3.060e-5 | 5.518e-4 | 2.234e-5 | 5.277e-4 | 2.549e-5 | 6.132e-4 | 8.864e-5 | 1.418e-3 |
| Diagonal anisotropic | 3.195e-5 | 5.490e-4 | 2.036e-5 | 4.569e-4 | 2.349e-5 | 5.438e-4 | 9.267e-5 | 1.358e-3 |
| Full anisotropic | 3.026e-5 | 5.518e-4 | 1.916e-5 | 4.024e-4 | 2.321e-5 | 5.169e-4 | 1.144e-4 | 1.703e-3 |

In order to compare the numerical errors of MRT-A, MRT-B1, MRT-B2 and MRT-B3 for varied dimensionless relaxation time $\tau$, the matrices $\mathbf{A}$ are given as

$$\mathbf{A}^{-1} = \begin{pmatrix} \tau & 0 \\ 0 & \tau \end{pmatrix}, \begin{pmatrix} \tau & 0 \\ 0 & 4\tau - 1.5 \end{pmatrix}, \begin{pmatrix} \tau & \tau - 0.5 \\ \tau - 0.5 & 4\tau - 1.5 \end{pmatrix}, \tag{44}$$

which represent the isotropic diffusion, diagonal anisotropic diffusion and full anisotropic diffusion problems, respectively; and the diffusion coefficient matrices are determined by $\mathbf{D} = \delta_t (\mathbf{I} - \mathbf{A}/2) \mathbf{A}^{-1} / 3$. Fig. 2 shows the average errors $Er_2$ for varied $\tau$ at time $t_m$ and $\mathbf{u} \equiv \mathbf{0}$ (pure diffusion problems) and $\mathbf{u} = (0.05, 0.05)$ (convection-diffusion problems). It can be seen that: (1) for pure diffusion problems, $Er_2$ of MRT-A are larger than those of MRT-B1, MRT-B2 and MRT-B3 generally, note that for diagonal anisotropic and full anisotropic diffusion problems (as shown by Fig. 2 (b) and (c)), the differences among $Er_2$ of MRT-A, MRT-B1, MRT-B2 and MRT-B3 are relatively small; (2) for convection-diffusion problems, $Er_2$ of MRT-A, MRT-B1 and MRT-B2 are all smaller than those of MRT-B3 obviously, note that although the deviation term disappears in MRT-A, MRT-B1 and MRT-B2 when $\mathbf{u} = const$, $Er_2$ of MRT-A are smaller than those of MRT-B1 and MRT-B2 for relatively large $\tau$ and are a little larger than those of MRT-B1 and MRT-B2 when $\tau$ is close to 0.5.

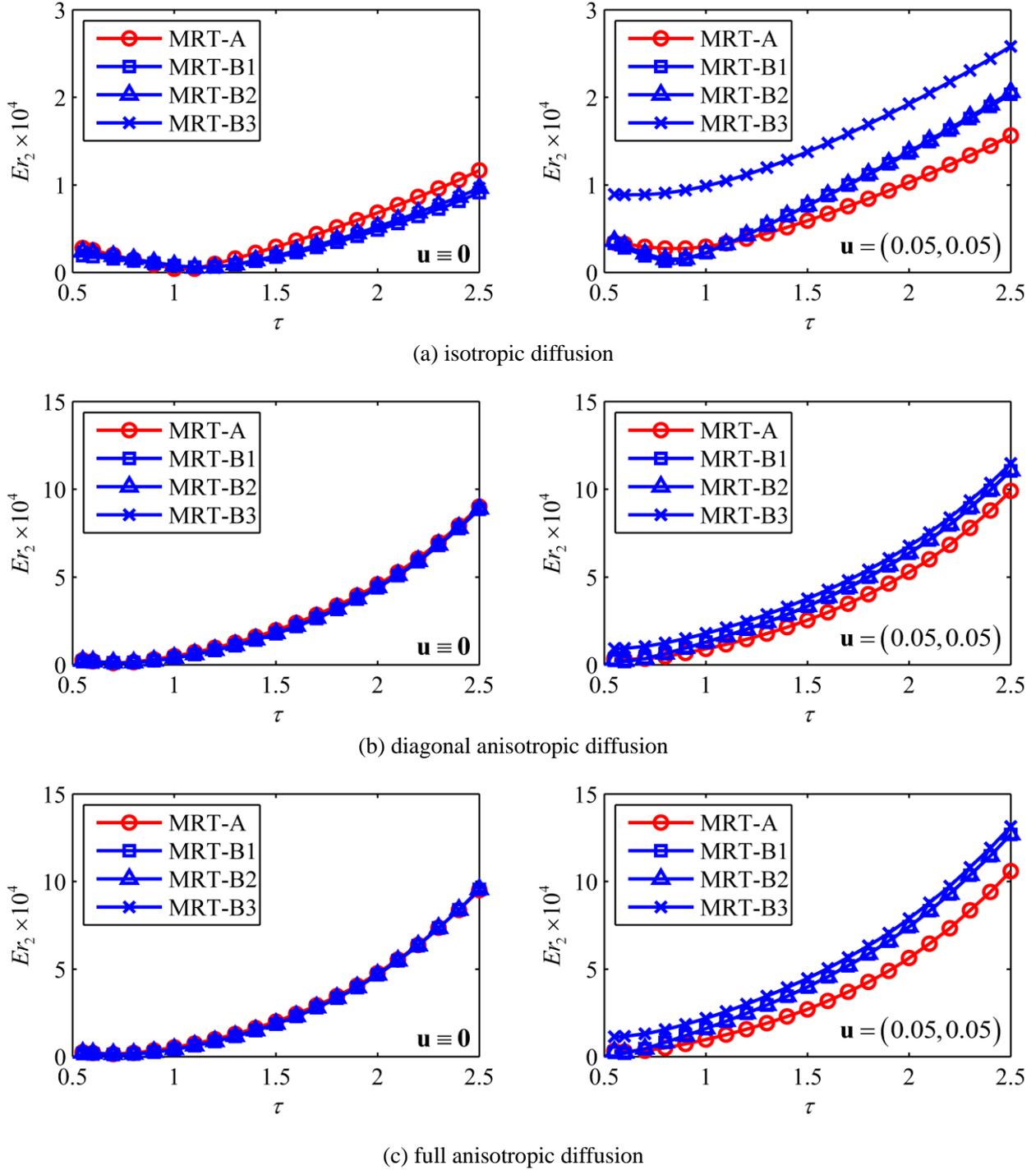

(a) isotropic diffusion

(b) diagonal anisotropic diffusion

(c) full anisotropic diffusion

**Fig. 2.** Variations of the average errors $Er_2$ of MRT-A, MRT-B1, MRT-B2 and MRT-B3 with the dimensionless relaxation times $\tau$ at time $t_m$ and $\mathbf{u} \equiv \mathbf{0}$ (pure diffusion problems) and $\mathbf{u} = (0.05, 0.05)$ (convection-diffusion problems) for (a) isotropic diffusion problems, (b) diagonal anisotropic diffusion problems, and (c) full anisotropic diffusion problems.

To investigate the convergence rate of the present modified MRT model (MRT-A), pure diffusion (i.e., $\mathbf{u} \equiv \mathbf{0}$) of a Gaussian hill is used here since $\phi$ at the boundary is quite small and the boundary effect on numerical result can be neglected. Considering the fact that the relaxation factors have an influence on numerical result, $\delta_x$ and $\delta_t$ are fixed at 1/256 when the lattice size $N$ in $x$ (or $y$) direction varies. So that the computation domain is $x \in [-L, L]$, $y \in [-L, L]$ where $L = \delta_x N/2$, and the initial variance is $\sigma_0^2 = (0.05L)^2$. The initial value of $\phi$ and the diffusion coefficient matrix $\mathbf{D}$ are also given by Eqs. (40) and (42), respectively. With the relation $\mathbf{D} = \delta_t (\mathbf{I} - \mathbf{A}/2) \mathbf{A}^{-1}/3$, matrix $\mathbf{A}$ is unvaried when $N$ changes and the other relaxation factors are set as $s_0 = 1.0$, $s_1 = s_2 = 1.1$, $s_4 = s_{xx}$, $s_6 = s_{yy}$, $s_7 = s_8 = 1.2$ consistently for all cases. The relative errors $Er$ defined as

$$Er = \frac{\sqrt{\sum_{x,y}(\phi_{numeric} - \phi_{analytic})^2}}{\sqrt{\sum_{x,y}\phi_{analytic}^2}} \tag{45}$$

are calculated here at time $t_m$. Note that the value of $t_m$ is changing when $N$ varies. The results are shown in Fig. 3. As seen, the present model for CDE (including isotropic, diagonal anisotropic and full anisotropic diffusion problems) has second order convergence rate. Note that although the deviation term is eliminated, the accuracy of the present model is still second order because the deviation term has an influence on accuracy (i.e., the magnitude of numerical error) rather than the order of accuracy [41, 42].

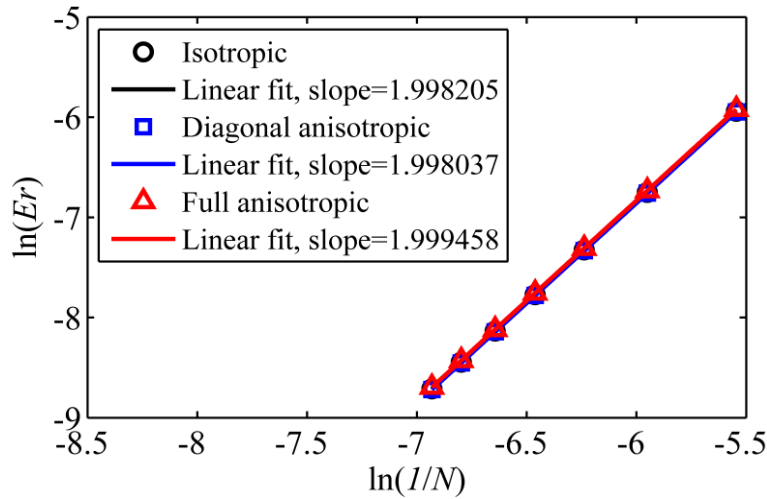

**Fig. 3.** Relative errors versus lattice sizes for pure diffusion of a Gaussian hill, where the black solid line is the linear fit of ln($Er$) *vs* ln(1/$N$) for isotropic diffusion problem, the blue solid line is the linear fit of ln($Er$) *vs* ln(1/$N$) for diagonal anisotropic diffusion problem, and the red solid line is the linear fit of ln($Er$) *vs* ln(1/$N$) for full anisotropic diffusion problem.

*3.2. Taylor-Aris dispersion*

In order to compare MRT-A with MRT-B1, MRT-B2 and MRT-B3 further, we consider the dispersion of the scalar variable $\phi$ under the parabolic background flow between two parallel plates of infinite length, which is known as the Taylor-Aris dispersion problem [43, 44], as shown in Fig. 4. The velocity profile is $u_x = 4Uy(W-y)/W^2$ for $0 \leq y \leq W$ and $u_y \equiv 0$, where $U$ is the maximum velocity, $W$ is the width of the channel. The diffusion coefficient matrix is $\mathbf{D} = \alpha\mathbf{I}$, which means the isotropic diffusion problem. After a long time, the average value of the scalar variable over the channel cross section, $\bar{\phi}$, tends to be a Gaussian distribution, no matter what the initial condition is. The longitudinal dispersion coefficient is defined as

$$\alpha^L = \frac{1}{2}\frac{d(\sigma^2)}{dt}, \tag{46}$$

where $\sigma^2$ is the variance of the Gaussian distribution calculated by

$$\sigma^2 = \frac{\int x^2 \bar{\phi}\,dx}{\int \bar{\phi}\,dx} - \left(\frac{\int x\bar{\phi}\,dx}{\int \bar{\phi}\,dx}\right)^2. \tag{47}$$

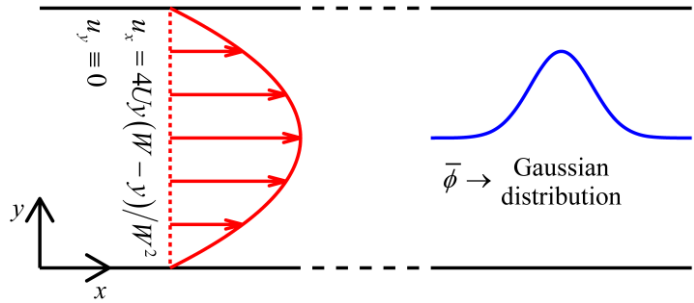

**Fig. 4.** Schematic of the Taylor-Aris dispersion between two parallel plates of infinite length with the coordinate, velocity profile and average value profile of the scalar variable after a long time shown.

In the numerical test, the channel width is $W = 1$ and divided into 32 lattices (note that the boundary grid nodes locate in the boundary), the length is $L = 150W$ and divided into 4800 lattices, the longitudinal dispersion coefficient $\alpha^L$ is observed during the evolution of an initial pulse at $x \in [0.197L, 0.203L]$. In the longitudinal direction, the periodic boundary condition is adopted; while on the bottom and top plates, the Neumann boundary condition with $\partial\phi/\partial y = 0$ is adopted. Note that this Neumann boundary condition can be realized easily by appointing that $f_2 = f_4$, $f_5 = f_8$ and $f_6 = f_7$ on the bottom plate; $f_4 = f_2$, $f_7 = f_6$ and $f_8 = f_5$ on the top plate. The relaxation factors are chosen as $s_0 = 1.0$, $s_1 = s_2 = 1.1$, $s_4 = s_6 = s_{xx} = s_{yy} = 1/\tau$, $s_{xy} = s_{yx} = 0$, $s_7 = s_8 = 1.2$ for all cases.

In order to compare the four models quantitatively, the relative errors, $Er$, of the longitudinal dispersion coefficient between the numerical results and the analytical solutions are calculated. The definition of $Er$ is

$$Er = \frac{\left|\alpha^L_{numeric} - \alpha^L_{analytic}\right|}{\alpha^L_{analytic}}, \tag{48}$$

where $\alpha^L_{numeric}$ is the convergent numerical result of $\alpha^L$ and satisfies the following convergence criterion

$$\frac{\alpha^L_{n+100} - \alpha^L_n}{\alpha^L_n} \leq 10^{-6}, \tag{49}$$

in which the subscript $n$ represents the $n$th time step; $\alpha^L_{analytic}$ is the analytical solution of $\alpha^L$ calculated by $\alpha^L_{analytic} = \alpha\left(1 + Pe^2/210\right)$ [33], here $Pe$ is the Péclet number defined as $Pe = 2UW/(3\alpha)$.

Table 3 gives the numerical results of MRT-A, MRT-B1, MRT-B2 and MRT-B3 at a fixed dimensionless relaxation time $\tau = 2$ when $Pe$ varies from 2 to 16. Note that the maximum velocity $U$ ($U = (\tau - 0.5)\delta_t Pe/2$) and the corresponding Mach number ($Ma \propto U$) increase with the increase of $Pe$ at a fixed dimensionless relaxation time. From Table 3, it can be found that: (1) as $Pe$ increases (i.e., $Ma$ increases), the relative errors of all models increase remarkably; (2) at a fixed $Pe$ (i.e., fixed $Ma$), the errors of MRT-A are always much smaller than those of MRT-B1, MRT-B2, and MRT-B3. Particularly, at $Pe = 16$, $Er$ of MRT-B1, MRT-B2 and MRT-B3 are 1.913595%, 1.949309% and 8.202396%, while $Er$ of MRT-A is only 0.503162%. Furthermore, as $Pe$ increases, the variation of the ratio among the relative errors of the four models is rather tiny for fixed dimensionless relaxation time ($Er_{MRT-A} : Er_{MRT-B1} : Er_{MRT-B2} : Er_{MRT-B3} \approx 1 : 3.787 : 3.858 : 16.302$ for $\tau = 2$). In order to verify this point further, Table 4 and Table 5 give the numerical results for the cases: (1) $\tau = 1$ and $Pe$ varies from 4 to 32, (2) $\tau = 0.55$ and $Pe$ varies from 16 to 128. It can be found that the ratio among the relative errors of the four models at fixed $\tau$ varies slightly as $Pe$ increases ($Er_{MRT-A} : Er_{MRT-B1} : Er_{MRT-B2} : Er_{MRT-B3} \approx 1 : 6.722 : 6.938 : 13.143$ for $\tau = 1$, $Er_{MRT-A} : Er_{MRT-B1} : Er_{MRT-B2} : Er_{MRT-B3} \approx 1 : 1.165 : 1.172 : 1.103$ for $\tau = 0.55$). This phenomenon suggests that the advantage of MRT-A can be maintained as $Ma$ increases (i.e., $U$ increases) if the dimensionless relaxation time $\tau$ keeps unvaried. Therefore, comparisons among the relative errors when $\tau$ varies will be carried out for fixed $U$ below (with $U$ keeping unvaried, varied $\tau$ is achieved by changing $Pe$).

Table 3

Comparisons of $Er$ among the numerical results of MRT-A, MRT-B1, MRT-B2 and MRT-B3 at $\tau = 2$.

| $Pe$ | $Er \times 100$ | | | | $Er_{\text{MRT-A}}:Er_{\text{MRT-B1}}:Er_{\text{MRT-B2}}:Er_{\text{MRT-B3}}$ |
| --- | --- | --- | --- | --- | --- |
| | MRT-A | MRT-B1 | MRT-B2 | MRT-B3 | |
| 2 | 0.017119 | 0.064699 | 0.065916 | 0.279083 | 1 : 3.779 : 3.850 : 16.303 |
| 4 | 0.064844 | 0.245131 | 0.249732 | 1.057056 | 1 : 3.780 : 3.851 : 16.302 |
| 8 | 0.213937 | 0.809726 | 0.824912 | 3.487515 | 1 : 3.785 : 3.856 : 16.302 |
| 16 | 0.503162 | 1.913595 | 1.949309 | 8.202396 | 1 : 3.803 : 3.874 : 16.302 |

Table 4

Comparisons of $Er$ among the numerical results of MRT-A, MRT-B1, MRT-B2 and MRT-B3 at $\tau = 1$.

| $Pe$ | $Er \times 100$ | | | | $Er_{\text{MRT-A}}:Er_{\text{MRT-B1}}:Er_{\text{MRT-B2}}:Er_{\text{MRT-B3}}$ |
| --- | --- | --- | --- | --- | --- |
| | MRT-A | MRT-B1 | MRT-B2 | MRT-B3 | |
| 4 | 0.007239 | 0.048551 | 0.050108 | 0.095078 | 1 : 6.707 : 6.922 : 13.134 |
| 8 | 0.023857 | 0.160225 | 0.165371 | 0.313692 | 1 : 6.716 : 6.932 : 13.149 |
| 16 | 0.056128 | 0.377198 | 0.389294 | 0.737770 | 1 : 6.720 : 6.936 : 13.144 |
| 32 | 0.084784 | 0.571907 | 0.590179 | 1.114430 | 1 : 6.745 : 6.961 : 13.144 |

Table 5

Comparisons of $Er$ among the numerical results of MRT-A, MRT-B1, MRT-B2 and MRT-B3 at $\tau = 0.55$.

| $Pe$ | $Er \times 100$ | | | | $Er_{\text{MRT-A}}:Er_{\text{MRT-B1}}:Er_{\text{MRT-B2}}:Er_{\text{MRT-B3}}$ |
| --- | --- | --- | --- | --- | --- |
| | MRT-A | MRT-B1 | MRT-B2 | MRT-B3 | |
| 16 | 0.163156 | 0.190019 | 0.191218 | 0.179961 | 1 : 1.165 : 1.172 : 1.103 |
| 32 | 0.246492 | 0.287021 | 0.288859 | 0.271816 | 1 : 1.164 : 1.172 : 1.103 |
| 64 | 0.282549 | 0.329003 | 0.331119 | 0.311598 | 1 : 1.164 : 1.172 : 1.103 |
| 128 | 0.293270 | 0.341531 | 0.343724 | 0.323424 | 1 : 1.165 : 1.172 : 1.103 |

Fig. 5 shows the relative errors of MRT-A, MRT-B1, MRT-B2 and MRT-B3 and dimensionless relaxation time at $U = 0.1$ when $Pe$ varies from 2 to 128. Note that at a fixed maximum velocity (i.e., at a fixed $Ma$), the dimensionless relaxation time $\tau$ ($\tau = 2U/(\delta_t Pe) + 0.5$) decreases with the increase of $Pe$. It can be seen from Fig. 5 that: (1) the relative errors of MRT-A are always smaller than those of MRT-B1 and MRT-B2, and the gap tends to become lager as $\tau$ increases; (2) the relationship between relative errors of MRT-A and MRT-B3 can be divided into three regions: (i) at a relatively large $\tau$ range ($\tau \geq 0.75$) with a corresponding smaller $Pe$ range ($Pe \leq 25.33$), the relative errors of MRT-A are smaller than those of MRT-B3, and the gap is also becoming lager as $\tau$ increases; (ii) at a medium $\tau$ range ($0.63 \leq \tau \leq 0.75$) with a corresponding medium $Pe$ range ($25.33 \leq Pe \leq 47.86$), the relative errors of MRT-B3 are a little smaller than those of MRT-A unexpectedly, which may be attributed to the neutralization between the numerical error induced by the deviation term and the other numerical errors of MRT-B3 in this range; (iii) at a relatively small $\tau$ range ($\tau \leq 0.63$) with a corresponding larger $Pe$ range ($Pe \geq 47.86$), the relative errors of MRT-A are a little smaller than those of MRT-B3, and both increase with the decrease of $\tau$. Note that, at a relatively large dimensionless relaxation time ($\tau \geq 0.86$), the numerical error of MRT-B3 can reach an obviously large value, while the numerical errors of MRT-B1 and MRT-B2 decrease a little; this is attributed to the proper nonlinear terms of velocity in the equilibrium functions (Eqs. (38) and (39)) [34, 35]. Moreover, when $\tau$ tends to 0.5, the numerical errors of the four models become closer, which may suggest that the numerical error induced by the deviation term becomes subordinate when $\tau$ is close to 0.5.

It should be also pointed out that MRT-A is a little less expensive than MRT-B1 and MRT-B2. This can be explained as follows: although the collision processes of $m_3$ and $m_5$ in MRT-A are more complicated (which increase the amount of computations), no nonlinear terms of velocity exist in the corresponding equilibrium function (which decrease the amount of computations). Because of the complicated collision processes of $m_3$ and $m_5$, MRT-A is a little more expensive than MRT-B3. However, numerical results show that these differences among MRT-A, MRT-B1, MRT-B2 and MRT-B3 are rather small, within ±1.5%.

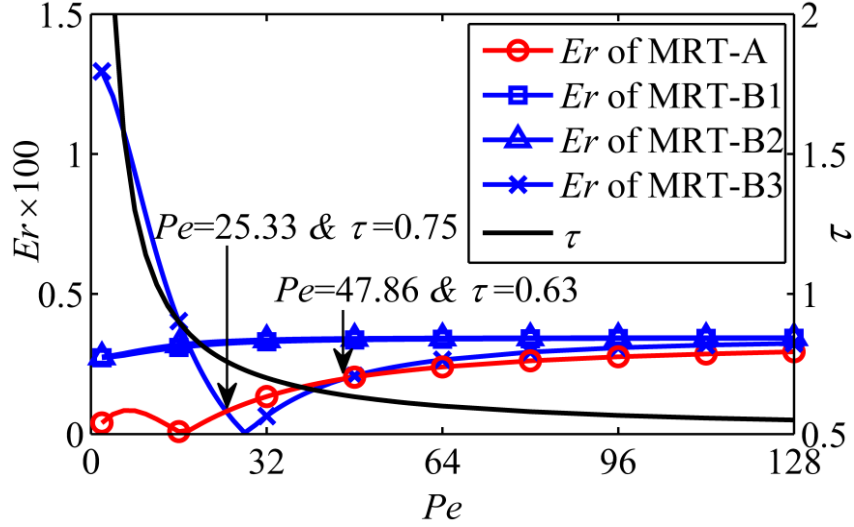

**Fig. 5.** Comparisons of the relative errors of MRT-A with MRT-B1, MRT-B2 and MRT-B3 at $U = 0.1$ for the Taylor-Aris dispersion problem with inconstant velocity vector distribution.

## 4. Conclusions

In this paper, a modified lattice Boltzmann model with multiple relaxation times is proposed for the convection-diffusion equation (CDE) with anisotropic diffusion coefficient. Different from the previous SRT and MRT lattice Boltzmann models for CDE, where there exists deviation term (see Eqs. (9) and (29)) in the corresponding macroscopic equation when the velocity vector varies generally with space or time through the Chapman-Enskog analysis, the present model can recover the CDE with anisotropic diffusion coefficient correctly by modifying the relaxation matrix (see Eq. (31)), as well as choosing the corresponding equilibrium distribution functions properly (see Eq. (15)). The modified MRT model is validated by the simulation of the diffusion of a Gaussian hill and the Taylor-Aris dispersion problems where the analytical solutions exist. From the numerical analysis and comparison, it can be found that the present modified MRT model can further reduce the numerical error under the condition of non-zero velocity vector, especially when the dimensionless relaxation time is relatively large.

## Acknowledgments

This work was supported by the National Natural Science Foundation of China through Grants No. 51376130 and No. 50925624, National Basic Research Program of China (973 Program) through Grant No. 2012CB720404, and Science and Technology Commission of Shanghai Municipality through Grant No. 12JC1405100.

## Appendix A. Three-dimensional version with D3Q15

In three-dimensional situations, D3Q15 discrete velocity model can be used (note that D3Q7 is not suitable for the present modified MRT model), and the discrete velocities are

$$\mathbf{e}_i = \begin{cases} c(0,0,0), & i=0, \\ c(\pm 1,0,0), c(0,\pm 1,0), c(0,0,\pm 1), & i=1\sim 6, \\ c(\pm 1,\pm 1,\pm 1), & i=7\sim 14. \end{cases} \tag{A.1}$$

When the lattice speed satisfies $c \equiv 1$, the linear transformation matrix can be chosen as [45]

$$\mathbf{M} = \begin{bmatrix} 1 & 1 & 1 & 1 & 1 & 1 & 1 & 1 & 1 & 1 & 1 & 1 & 1 & 1 & 1 \\ -2 & -1 & -1 & -1 & -1 & -1 & -1 & 1 & 1 & 1 & 1 & 1 & 1 & 1 & 1 \\ 16 & -4 & -4 & -4 & -4 & -4 & -4 & 1 & 1 & 1 & 1 & 1 & 1 & 1 & 1 \\ 0 & 1 & -1 & 0 & 0 & 0 & 0 & 1 & -1 & 1 & -1 & 1 & -1 & 1 & -1 \\ 0 & -4 & 4 & 0 & 0 & 0 & 0 & 1 & -1 & 1 & -1 & 1 & -1 & 1 & -1 \\ 0 & 0 & 0 & 1 & -1 & 0 & 0 & 1 & 1 & -1 & -1 & 1 & 1 & -1 & -1 \\ 0 & 0 & 0 & -4 & 4 & 0 & 0 & 1 & 1 & -1 & -1 & 1 & 1 & -1 & -1 \\ 0 & 0 & 0 & 0 & 0 & 1 & -1 & 1 & 1 & 1 & 1 & -1 & -1 & -1 & -1 \\ 0 & 0 & 0 & 0 & 0 & -4 & 4 & 1 & 1 & 1 & 1 & -1 & -1 & -1 & -1 \\ 0 & 2 & 2 & -1 & -1 & -1 & -1 & 0 & 0 & 0 & 0 & 0 & 0 & 0 & 0 \\ 0 & 0 & 0 & 1 & 1 & -1 & -1 & 0 & 0 & 0 & 0 & 0 & 0 & 0 & 0 \\ 0 & 0 & 0 & 0 & 0 & 0 & 0 & 1 & -1 & -1 & 1 & 1 & -1 & -1 & 1 \\ 0 & 0 & 0 & 0 & 0 & 0 & 0 & 1 & 1 & -1 & -1 & -1 & -1 & 1 & 1 \\ 0 & 0 & 0 & 0 & 0 & 0 & 0 & 1 & -1 & 1 & -1 & -1 & 1 & -1 & 1 \\ 0 & 0 & 0 & 0 & 0 & 0 & 0 & 1 & -1 & -1 & 1 & -1 & 1 & 1 & -1 \end{bmatrix}. \tag{A.2}$$

The equilibrium function is chosen as

$$m_0^{eq} = m_0 = \phi, \tag{A.3a}$$

$$m_1^{eq} = \frac{5}{6}\alpha_1\phi, \tag{A.3b}$$

$$m_2^{eq} = \frac{1}{24}\alpha_2\phi, \tag{A.3c}$$

$$m_3^{eq} = c_1\phi u_x, \qquad m_5^{eq} = c_1\phi u_y, \qquad m_7^{eq} = c_1\phi u_z, \tag{A.3d}$$

$$m_4^{eq} = \frac{1}{4}c_2\phi u_x, \qquad m_6^{eq} = \frac{1}{4}c_2\phi u_y, \qquad m_8^{eq} = \frac{1}{4}c_2\phi u_z, \tag{A.3e}$$

$$m_9^{eq} = m_{10}^{eq} = m_{11}^{eq} = m_{12}^{eq} = m_{13}^{eq} = m_{14}^{eq} = 0. \tag{A.3f}$$

The parameters in the equilibrium function meet the following conditions

$$c_1 = 1, \quad c_1 + c_2/4 = 0. \tag{A.4}$$

The relaxation matrix, **S**, is described as

$$\mathbf{S} = \begin{bmatrix}
s_0 & 0 & 0 & 0 & 0 & 0 & 0 & 0 & 0 & 0 & 0 & 0 & 0 & 0 & 0 \\
0 & s_1 & 0 & 0 & 0 & 0 & 0 & 0 & 0 & 0 & 0 & 0 & 0 & 0 & 0 \\
0 & 0 & s_2 & 0 & 0 & 0 & 0 & 0 & 0 & 0 & 0 & 0 & 0 & 0 & 0 \\
0 & 0 & 0 & s_{xx} & \left(\frac{s_{xx}}{2}-1\right)s_4 & s_{xy} & \frac{s_{xy}}{2}s_6 & s_{xz} & \frac{s_{xz}}{2}s_8 & 0 & 0 & 0 & 0 & 0 & 0 \\
0 & 0 & 0 & 0 & s_4 & 0 & 0 & 0 & 0 & 0 & 0 & 0 & 0 & 0 & 0 \\
0 & 0 & 0 & s_{yx} & \frac{s_{yx}}{2}s_4 & s_{yy} & \left(\frac{s_{yy}}{2}-1\right)s_6 & s_{yz} & \frac{s_{yz}}{2}s_8 & 0 & 0 & 0 & 0 & 0 & 0 \\
0 & 0 & 0 & 0 & 0 & 0 & s_6 & 0 & 0 & 0 & 0 & 0 & 0 & 0 & 0 \\
0 & 0 & 0 & s_{zx} & \frac{s_{zx}}{2}s_4 & s_{zy} & \frac{s_{zy}}{2}s_6 & s_{zz} & \left(\frac{s_{zz}}{2}-1\right)s_8 & 0 & 0 & 0 & 0 & 0 & 0 \\
0 & 0 & 0 & 0 & 0 & 0 & 0 & 0 & s_8 & 0 & 0 & 0 & 0 & 0 & 0 \\
0 & 0 & 0 & 0 & 0 & 0 & 0 & 0 & 0 & s_9 & 0 & 0 & 0 & 0 & 0 \\
0 & 0 & 0 & 0 & 0 & 0 & 0 & 0 & 0 & 0 & s_{10} & 0 & 0 & 0 & 0 \\
0 & 0 & 0 & 0 & 0 & 0 & 0 & 0 & 0 & 0 & 0 & s_{11} & 0 & 0 & 0 \\
0 & 0 & 0 & 0 & 0 & 0 & 0 & 0 & 0 & 0 & 0 & 0 & s_{12} & 0 & 0 \\
0 & 0 & 0 & 0 & 0 & 0 & 0 & 0 & 0 & 0 & 0 & 0 & 0 & s_{13} & 0 \\
0 & 0 & 0 & 0 & 0 & 0 & 0 & 0 & 0 & 0 & 0 & 0 & 0 & 0 & s_{14}
\end{bmatrix}. \tag{A.5}$$

Introducing $\mathbf{A} = \begin{pmatrix} s_{xx} & s_{xy} & s_{xz} \\ s_{yx} & s_{yy} & s_{yz} \\ s_{zx} & s_{zy} & s_{zz} \end{pmatrix}$ and through the Chapman-Enskog analysis, the corresponding macroscopic equation is

$$\partial_t \phi + \nabla \cdot (\phi \mathbf{u}) = \nabla \cdot \left[ \left( \frac{2}{3} + \frac{55}{54}\alpha_1 + \frac{1}{216}\alpha_2 \right) \delta_t \left( \mathbf{I} - \frac{\mathbf{A}}{2} \right) \mathbf{A}^{-1} \nabla \phi \right]. \tag{A.6}$$

The diffusion coefficient matrix is given by $\mathbf{D} = \left( \frac{2}{3} + \frac{55}{54}\alpha_1 + \frac{1}{216}\alpha_2 \right) \delta_t \left( \mathbf{I} - \frac{\mathbf{A}}{2} \right) \mathbf{A}^{-1}$, which can describe the anisotropic diffusion problems. As compared with the convection-diffusion equation, no deviation term emerges in the corresponding macroscopic equation.